\documentstyle[11pt,gh2001-asp,twoside,epsf]{article}
\begin{document}
\title{Observations of Disk Galaxy Evolution}
\author{Roberto G. Abraham}
\affil{Dept. of Astronomy \& Astrophysics, University of Toronto\\ 60 St. George Street, Toronto ON, M5S 3H8 Canada}
\author{Sidney van den Bergh}
\affil{Dominion Astrophysical Observatory, Herzberg Institute of Astrophysics\\National Research Council of Canada, 5071 W. Saanich Road, Victoria B.C., V9E 2E7 Canada}

\begin{abstract}
The morphologies of disk galaxies begin to deviate systematically from those of nearby galaxies at surprisingly low redshifts, possibly as low as $z=0.3$. This corresponds to a time $\sim3.5$ Gyr in the past, which is only one quarter of the present age of the Universe. Beyond $z = 0.5$ (a look-back time of 5 Gyr) the effects of evolution on spiral structure are rather obvious: spiral arms are less well-developed and more chaotic, and barred spiral galaxies seem rarer.  By $z=1$, around 30\% of the galaxy population is sufficiently peculiar that classification on Hubble's traditional tuning fork system is meaningless. On the other hand, the co-moving space density and the sizes of luminous disks have not changed significantly since $z=1$. Tully-Fisher measurements indicate that the dynamical state and luminosities of large disk systems are also consistent with passive evolution out to a redshift of unity. We conclude that the general appearance of luminous disk galaxies has continuously changed with cosmic epoch, but their overall numbers have been conserved since $z=1$, and the bulk of the stars in these systems may have formed at $z>1$. 
\end{abstract}

\section{Scope of This Review}

This review is an attempt to summarize current observational constraints on evolution in disk galaxies. Because of space limitations, we will mostly touch upon recent highlights, and do so in a simplified way that treats morphological, luminosity/size, and dynamical evolution independently (even though they are probably deeply interlinked). Throughout this review we will espouse the philosophy that the best way to understand  how galaxies evolve is to watch them evolve {\em in situ} at high redshifts. Since our remit is to focus on evolution in disks, it is necessary to direct most of our attention to observations capable of distinguishing between disk galaxies and other objects in the field at high redshifts.  We therefore unapologetically emphasize a morphological point of view,  and focus mostly on recent work undertaken with the {\em Hubble Space Telescope} (HST).

\section{A User's Guide to Interpreting High-z Disk Galaxy Observations}

\begin{figure}
%\plotfiddle{EPSFILE}{VSIZE}{ROT}{HSF}{VSF}{HTRANS}{VTRANS}
\plotfiddle{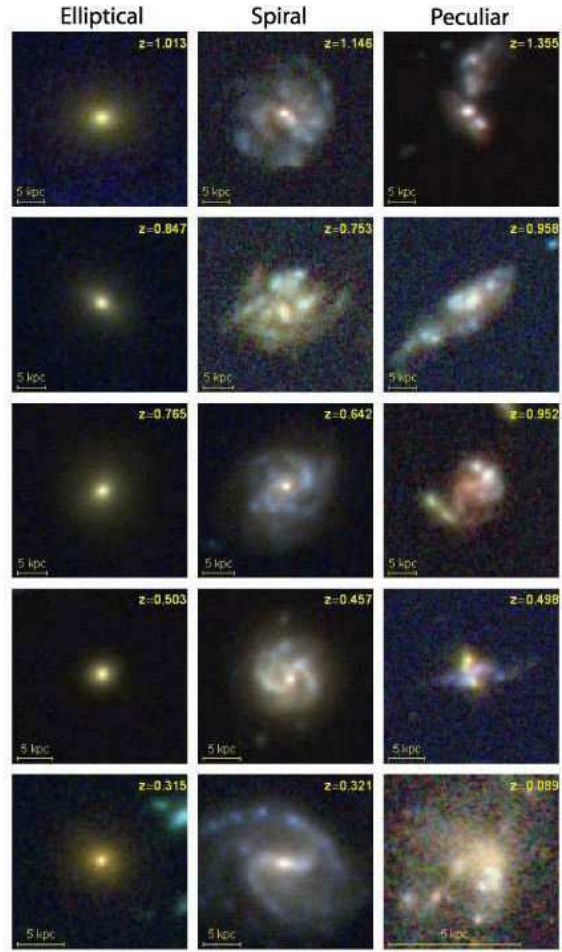} {12cm} {0} {50} {50} {-100} {0}
\caption{
Montage of typical elliptical (left column), spiral (middle column), and peculiar (right column) galaxies in the Northern and Southern Hubble Deep Fields, imaged by the Hubble Space Telescope. The images are sorted by redshift (local galaxies at bottom, distant galaxies at top).  Color images were constructed by stacking images obtained through blue, yellow, and near-infrared filters. Note the gradual loss in the organization (and the increase in the fragmentation) of spiral arms in the spirals. Barred spirals (such as the example shown in bottom row) become rare beyond z=0.5. The physical nature of most of the peculiar galaxies in the far right column is poorly understood. The exception is the object at the bottom-right corner, which is a dwarf irregular. 
\label{fig:montage}}
\end{figure}

Present observations of galaxy evolution extend to redshifts of at least $z\sim4$ (Steidel et al. 2001), and possibly much higher (Dickinson et al. 2000). This can be something of a double-edged sword when it comes to the interpretation of observations.  The tremendous benefits which come from being able to observe evolution directly at high redshifts come at the price of an extra layer of complexity in the interpretation of these observations, since it necessary to  disentangle true evolution from a myriad of possible observational biases. These biases ricochet through all studies of high-$z$ galaxy evolution, and how they are handled conditions the interpretation of a large body of work. For example, consider Figure~\ref{fig:montage}, which illustrates many of the evolving generic characteristics of high-redshift disks. How much of the spectacular evolution in the appearance of the typical disk systems as a function of redshift shown this figure is due to observational selection effects? 

We now know enough about high-$z$ galaxy morphology to let us really come to grips with this question. In order of increasing severity, the three effects which most impact high-redshift galaxy observations are:

\begin{enumerate}
\item {\bf  Resolution/pixellation effects.} In most cases these are only a minor source of bias. HST's resolving power and relatively small pixels, coupled with the inflection point in angular diameter distance as a function of redshift, mean than in many cases HST images actually have {\em higher} resolution than do calibration images of nearby  galaxies. For example, under typical seeing conditions, only those galaxies at $z\la0.05$ in the Sloan Digital Sky Survey have resolution equal to or better than that of HST at $z=1$. 
\item {\bf Bandshifting of the rest wavelength of observation.} Since most morphological studies of high-redshift galaxies have been done using HST's F814W filter (approximately $I$-band), these studies correspond to rest-frame $B$-band (where local morphology is best understood) at $z\sim0.7$. Therefore bandshifting effects are probably quite innocuous at $z\la1$,  but grow more worrisome at higher redshifts. As described below, there is already ample evidence for strong evolutionary effects by $z=0.7$, ruling out the extreme possibility (considered seriously early on) that most of the morphological changes witnessed on HST images are due to bandshifting effects. Furthermore, even the limited amount of imaging made with the HST Near-Infrared Camera Multi-Object Spectrometer (NICMOS) shows quite clearly that most $z>1$ objects classified as peculiar in $I$-band remain peculiar when imaged at rest optical wavelengths (Dickinson 1999). Finally, it is worth mentioning that the effects of bandshifting can be calibrated out of imaging data using pixel-by-pixel K-corrections provided deep data are available in several filters so that an estimate of an individual pixel's color can be made and a template spectrum assigned (Abraham 1997; Bouwens, Broadhurst \& Silk 1998).
\item {\bf The interplay between cosmological surface brightness dimming and evolution in stellar populations.}  This is certainly the most poorly constrained effect. Cosmological surface brightness diminution goes as $(1+z)^{-3}$ (Tolman dimming), but this dimming in apparent surface brightness is partially offset by an (uncertain) evolutionary brightening in intrinsic surface brightness which operates in the opposite direction. As we will now show, evolutionary synthesis predictions can go some way toward making predictions for higher redshifts.
\end{enumerate}

\begin{figure}
\plotone{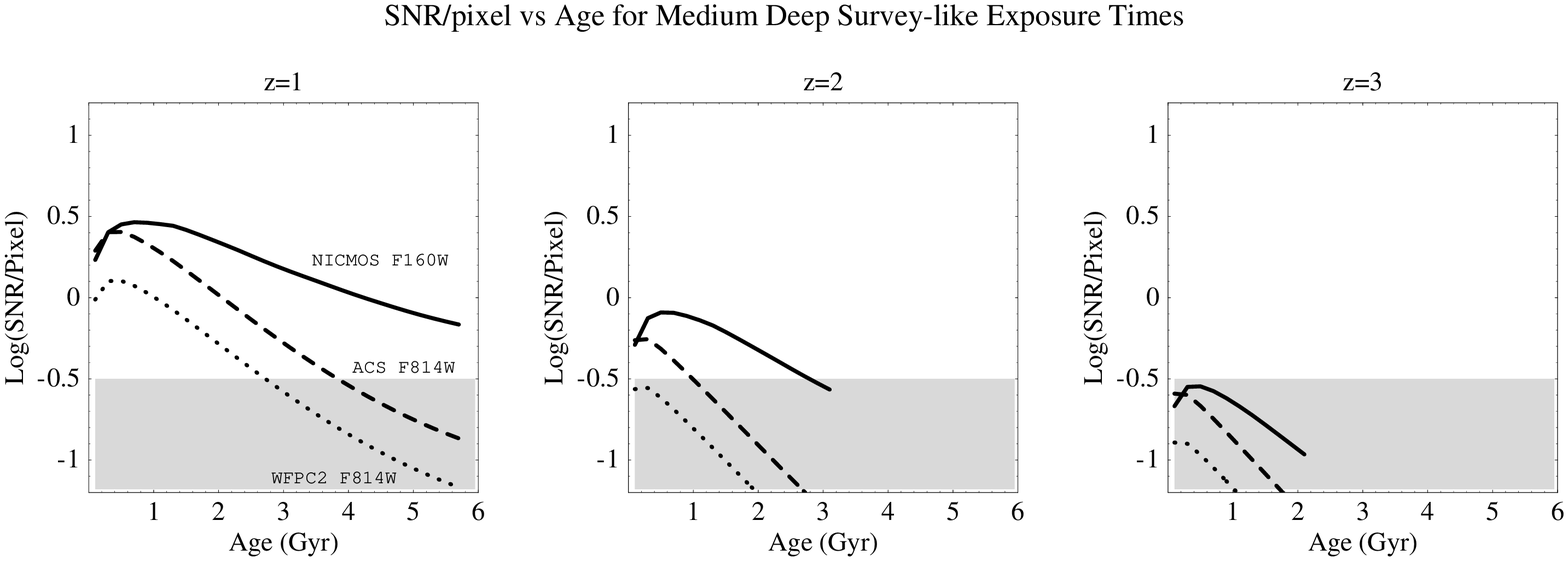}
\caption{
Spectral synthesis predictions for signal-to-noise/pixel (the appropriate figure of merit for work intended to probe resolved structures in galaxies) as a function of age for a 15ks ($\sim5$ orbit) exposure of a solar-neighborhood projected mass density seen at z=1 (left panel), z=2 (middle panel), and z=3 (right panel). Grey regions in this plot are unobservable due to low signal-to-noise. Another limit to accessible parameter space is the age of the Universe: curves shown are limited to ages less than that of the Universe at each epoch of observation. Tracks for various telescope/instrument combinations are shown. Dotted curves correspond to HST F814W imaging with the WFPC2 camera,  dashed curves correspond to HST $z'$ imaging with the Advanced Camera for Surveys, and solid curves correspond to HST F160W imaging with NICMOS. Calculations are based on a stellar population with an exponential star-formation history with an e-folding timescale of 1 Gyr. A Miller-Scalo IMF and 30\% solar metallicity is assumed, as well as a common rebinned pixel scale of 0.1\arcsec/pixel for each instrument.
\label{fig:sbVsage}}
\end{figure}

\begin{figure}
\plotone{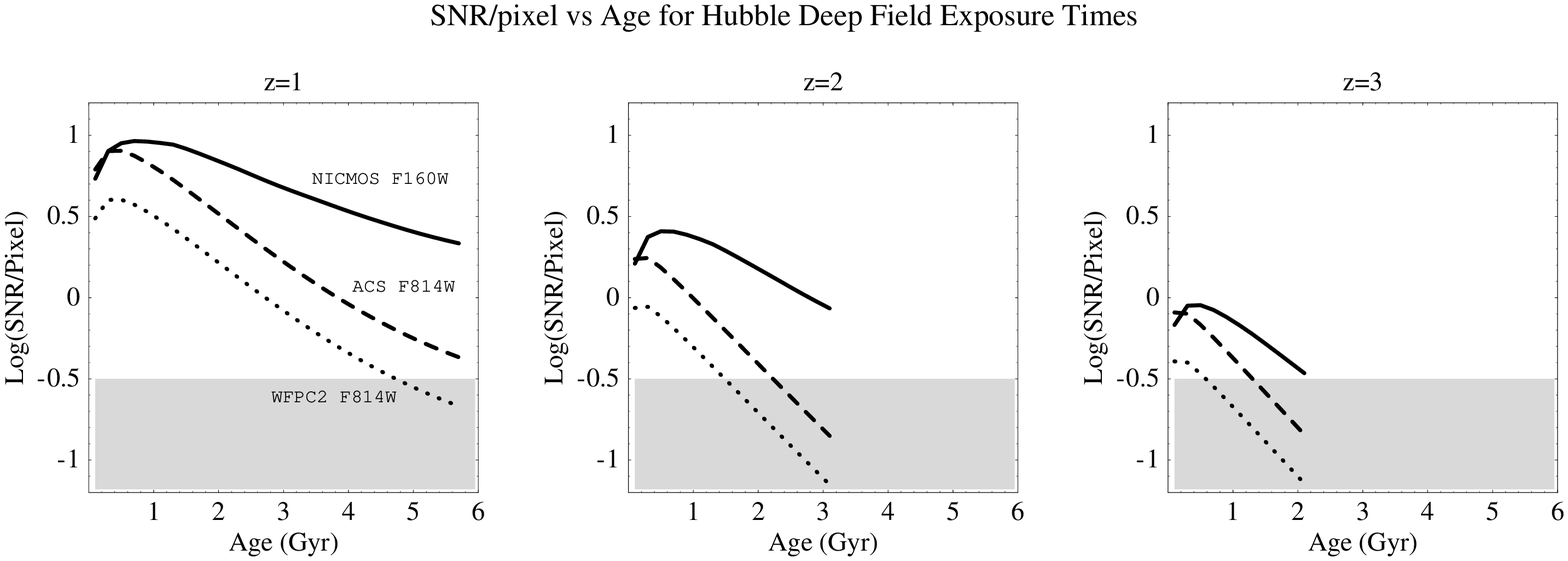}
\caption{
As for the previous figure, except assuming an exposure time of 150ks (approximately that of the Hubble Deep Field F814W-band observations).
\label{fig:sbVsageHDF}}
\end{figure}

 Figures~\ref{fig:sbVsage} and~\ref{fig:sbVsageHDF} are an attempt to come to grips simultaneously with the effects of bandshifting, cosmological surface brightness dimming, and evolution. These figures plot the signal-to-noise/pixel as a function of age and redshift for a projected stellar mass density corresponding to that of the Milky Way at the solar radius. The region shown in gray is invisible even after a modest amount of binning. We assume a stellar population with an exponentially declining star-formation history characterized by an e-folding timescale of 1~Gyr\footnote{Note that when modeling individual pixels in resolved galaxies, a single short-timescale exponential can be used to provide a quite general description of star-formation characteristics. Longer exponential timescale (or even constant) star-formation models used to model the integrated properties of disks are well-approximated by the spatial superposition of multiple resolved short timescale exponentials with a range of ages. }. In Figure~\ref{fig:sbVsage} we model 15ks ({\em i.e.} approximately five orbit) HST integrations typical of moderately deep HST imaging in the Medium Deep Survey (Griffiths et al. 1994), the Hawaii Deep Survey (Cowie, Hu \& Songaila 1995), and the LDSS/CFRS (Brinchmann et al. 1998) survey. We use the known characteristics of the Wide-Field Planetary Camera 2 (WFPC2) in the F814W filter, as well as  the projected characteristics of the Advanced Camera for Surveys (ACS) in the SDSS $z'$ band, and the refurbished NICMOS in F160W (approximate $H$-band)\footnote{Both ACS and NICMOS are scheduled for installation aboard HST during Servicing Mission 3, currently scheduled for Feb/March 2002.}. The bottom line is that Figure~\ref{fig:sbVsage} shows that at $z\la1$, there is relatively little bias in optical observations undertaken with HST. At this redshift, projected mass densities corresponding the intermediate regions of large disk galaxies are observable for around 70\% of the maximum possible age for the galaxy (defined as the age of the Universe at the epoch of observation). However, the second panel in the figure shows that by $z=2$ biases with WFPC2 are severe: structural information at optical wavelengths comes only from the densest regions in galaxies (such as bulges), or from the youngest parts of galaxies (such as bright star-formation complexes with ages of less 1 Gyr). By $z=2$ it is simply no longer possible to use WFPC2 optical data to compare fairly fundamental characteristics, such as the size and morphology of galaxies, with those of local objects, at least without much deeper imaging than is possible with 5 orbits on HST. It is interesting to note that NICMOS allows fair mass densities to be probed at $z=2$, albeit at the cost of limited field of view. 

What is the corresponding situation for the much deeper {\em Hubble Deep Field} (HDF) observations? Figure~\ref{fig:sbVsage} shows that imaging at HDF-like depths with WFPC2 probes solar-neighborhood-like masses out to $z=2$ about as well as 5-orbit HST imaging does at $z=1$. At these depths ACS allows somewhat unbiased observations to be made out to $z=3$, while NICMOS is completely unbiased at this redshift. 

Armed with this machinery, we can answer with some confidence the rhetorical question posed at the beginning of this section. Because the systems shown in Figure~\ref{fig:montage} are seen at $z\la1$ and these images are from the HDF, it is clear that their morphological peculiarities are intrinsic, and thus a fair reflection of the underlying state of their stellar populations. If these systems had been at somewhat higher redshifts, but still at $z\la2$, the interpretation of the structural changes would depend on whether or not they were consistent with a simple relative brightening (or fading) of known morphological subcomponents. This might be a difficult determination to make in some cases, but at least we could be reasonably confident that most of the galaxy was not hidden below the surface brightness detection threshold, which might easily be the case at higher redshifts.

\section{Luminosity/Size Evolution}

Because most studies of luminosity and size evolution are based on diagnostic diagrams showing  central surface brightness plotted as a function of scale size, distinguishing between pure luminosity evolution and pure size evolution is difficult. For simplicity I will treat these two effects together. 

Based on ground-based imaging obtained with the CFHT, Schade et al. (1996a) found that galaxy disks at $0.5 < z < 1.1$ have a mean rest-frame, inclination-corrected central surface brightness of ${\mu(B)}_{AB} = 19.8 \pm 0.1$ mag arcsec$^{-2}$, which is $\sim 1.6$ mag brighter than the Freeman (1970) value. At low redshifts ($0.2 < z < 0.5$) these authors found a mean surface brightness of ${\mu(B)}_{AB} = 21.3 \pm 0.25$  mag arcsec$^{-2}$, which is consistent with the Freeman value. Assuming most of this effect is due to luminosity (rather than size) evolution, the broad-brush picture painted is one of $B$-band evolution proceeding roughly linearly (in magnitude units) with redshift. This remains the most popular prescription for the luminosity evolution of disks, and is quite consistent with most recent studies, the most glaring exception being Simard et al. (1999), as described below.  

The environmental dependance of disk surface brightness evolution has been explored in another ground-based sample studied by Schade et al. (1996b). These authors studied the differential properties of cluster and field galaxies from the CNOC survey. At redshifts of (0.23, 0.43, 0.55) the disk surface brightness in cluster late-type galaxies is higher in the $B$ band by $\Delta\mu(B)_0 = (-0.58 \pm 0.12, -1.22 \pm 0.17, -0.97 \pm  0.2)$mag, respectively, relative to Freeman's law, which is consistent with the disk brightening seen in the field. However, Schade et al. emphasize that their sample of cluster galaxies is dominated by objects at large distances (up to 3 Mpc) from the dense cluster cores, so the implications of these findings for studies of the Butcher-Oemler effect and the morphology-density relation really must await investigations that thoroughly probe galaxy properties as a function of clustercentric radius. 

The best HST-based follow-up investigations of luminosity/size evolution are the studies by Lilly et al. (1998) and Simard et al. (1999), which are based on samples with known redshifts obtained from the LDSS/CFRS and DEEP surveys, respectively. Lilly et al. (1999) examined the  two-dimensional surface brightness profiles for 341 objects at  $0 < z < 1.3$. The size function of disk scale lengths in disk-dominated galaxies was found to stay roughly constant to $z \sim 1$, at least for large disks where the sample was most complete and where the disk and bulge decompositions were most reliable. Lilly et al. define these objects to be those with exponential scale lengths $> 3.2 h^{-1}_{50}$ kpc, and find good agreement between the scale lengths measured in their sample and a local calibration sample. This important result suggests that typical disks cannot have grown substantially with cosmic epoch since $z=1$, unless this growth is balanced by a corresponding destruction in the number of large disks through merging. It is probably worth emphasizing that while the disks in this sample appear to be the same size at high-$z$,  they show significant differences from local disks: they have higher $B$-band disk surface brightnesses, bluer overall $(U-V)$ colors, higher [O II] equivalent widths, and less regular morphologies than local counterparts.  Lilly et al. conclude this is consistent with an increase in the star formation rate by a factor of about three to $z \sim 0.7$, in agreement with the expectations from recent models for the evolution of the Milky Way disk.

The measurements presented by Simard et al. (1999) are seemingly in  close agreement with those given by Lilly et al. (1998), but these authors differ greatly in their interpretation. The Simard et al. study is based on a total sample of 190 field galaxies ($I_{814}\le23.5$ mag and $0.1<z<1.1$), and these authors find that if selection effects are ignored the mean disk surface brightness (averaged over all galaxies) increases by $\sim1.3$ mag from z=0.1 to 0.9, in fairly good agreement with the Lilly et al. (1998) and Schade et al. (1995a,b) results. However, by creating mock catalogs and simulating the appearance of galaxies, these authors are able to quantify the surface brightness selection function for the sample as a whole, and  conclude that  most of this change could be due to comparing low-luminosity galaxies in nearby redshift bins to high-luminosity galaxies in distant bins. After allowing for this effect, these authors find that no discernible evolution remains in the disk surface brightness of bright ($M_B<-19$) disk-dominated galaxies, and that at all redshifts  to $z\sim 1$ the magnitude-size envelope for the bulk of galaxies is consistent with that of local galaxies. While the derivation of a formal selection function for the dataset based on simulations is laudable, we remain suspicious of the final result because the accessible area of size-surface brightness parameter space in a deep survey would seem to be larger than the region of parameter space occupied by real galaxies. For example, Driver (1999) uses the Hubble Deep Field to derive the ``local'' ($0.3<z<0.5$) bivariate brightness distribution of field galaxies within a modest (326 Mpc$^3$, 47 galaxies) {\em volume-limited} sample. The sample probes the underlying galaxy population within the limits $0.3<z<0.5$, $-21.3<M_B<-13.7$ mag, $18.0<\mu_B<24.55$ mag arcsec$^{-2}$. Galaxies occupy a relatively narrow locus in this parameter space, and Driver concludes that luminous low-surface brightness galaxies account for less than 10\% of the $L^\star$  population. We conclude that it is unlikely that surface brightness selection plays a significant role in conditioning the numbers for photometric disk evolution, and that ``what you see is what you get'' when interpreting the size-magnitude diagram in the present generation of deep surveys which reach $L^\star$ at $z\la1$.

Photometric redshifts can also be employed to constrain the sizes of high-redshift disk galaxies at higher redshifts than is possible spectroscopically, though with correspondingly larger uncertainties. In an interesting paper, Giallongo et al. (2000) compare the intrinsic sizes of the field galaxies with $I<26 $ mag in the Hubble Deep Field and the NTT Deep Field to the predictions from a Cold Dark Matter (CDM) model, finding that  the distribution of the galaxy disk sizes at different cosmic epochs is within the range predicted by typical CDM models, modulo an excess of small-size disks at $z>0.5$. This result is similar to earlier results obtained by Bouwens et al. (1998a,b) who adopted the novel (and, in our view, excellent) technique of ``cloning'' foreground galaxies to model their high-redshift counterparts in order to determine that faint HDF galaxies ($I > 24$ mag) are much smaller, more numerous, and less regular than predicted by no-evolution models. These authors find that pure luminosity evolution in a flat $\Lambda$-dominated cosmology can account for the properties of the brightest and biggest galaxies in the HDF, but that the excess of small faint galaxies remains difficult to explain.  In any case, when speculating on the nature of the small, faint galaxy population on HST images we caution that it may not be too meaningful to try to understand such systems in the context of local well-ordered disks.  As described in the next two sections, disk warping and morphological peculiarity are seen to become more important with increasing redshift, which makes one suspicious that the notion of an isolated well-ordered disk may disappear at $z\ga1.5$.

\section{Dynamical Evolution}

Dynamical investigations of disk galaxy evolution are in their infancy. Ambitious programs (such as the DEEP project) are in progress, but the published results from these comprehensive investigations have yet to appear. Vogt et al. (1997) reports initial results from a Keck-based study of the internal kinematics of eight objects (spanning the range $-21.8 \la M_B \la -19.0$ mag for $H_0 = 75$ km s$^{-1}$ Mpc$^{-1}$ and $q_0 = 0.05$, i.e. predeominantly sub-$L^\star$).  A small offset of $\la0.4$ mag in $B$-band is seen, but no obvious change in the shape or slope of the relation with respect to the local Tully-Fisher relation is detected. The offset seems entirely consistent with the passive evolution described in the previous section. The DEEP project has reported obtaining kinematical information for over 100 high-redshift galaxies, and the preliminary results from this study have already been reported several times at meetings (Vogt 2000, 2001; Faber et al 2001). These data form a high redshift Tully-Fisher relation that spans five magnitudes and extends to well below $L^\star$, and the basic result reported in Vogt et al. (1997) seems unchanged: no obvious change in shape or slope relative to the local Tully-Fisher relation is seen, while a smooth but modest increase in disk surface brightness as a function of redshift is detected. 

Since there appears to be strong kinematical evidence for luminous, large disk systems at high redshifts, it is interesting to enquire as to the highest redshift at which a bona-fide luminous disk galaxy has been seen. The current record holder is the object reported by van Dokkum and Sanford (2001), who discovered a rapidly rotating disk  galaxy at $z=1.34$ with maximum rotation velocity $V_{\rm max}=290\pm20$ km $s^{-1}$, on the basis of resolved [OII] emission.  The galaxy has $M_V=-22.4\pm0.2$ ($\sim3L^\star$) in the rest frame, and its spectral energy distribution is very well fitted by that of a redshifted Sb/c galaxy.  The observed kinematics, morphology (both HST optical and ground-based infrared), and spectral energy distribution are consistent with a massive spiral galaxy. Interestingly, the lower limit on $V_{\rm max}$ gives an upper limit on the offset from the present-day Tully-Fisher relation, and the galaxy is overluminous by less than $0.7\pm0.4$ mag in rest-frame $V$-band. Of course, when it becomes possible (in the not-too-distant future) to construct samples of $z>1$ galaxies to probe the high-$z$ Tully-Fisher relation one will need to be very careful, since if disks are in their infancy at these redshifts they may be so tilted and chaotic that it is not clear what the physical nature of the ``rotation velocity"  really is. Resolved spectroscopy with Integral Field Units (see the paper by Bershady in this volume) may be essential for the success of dynamical programs at high-$z$.

\section{Morphological Evolution}

The number counts of disk galaxies are in good agreement with the predictions of passive-evolution models to at least $I_{AB}=25.5$ mag, the effective limit for crude morphological classification (into very broad early/disk/peculiar bins)  in the Hubble Deep Fields. At this magnitude limit around 30\% of all galaxies in the field are classified as ``peculiar/merging", {\em i.e.}  they fall outside the conventional framework of the Hubble System altogether. As of this writing this basic result has been derived independently using at least three different methods for automatic classification (Abraham et al. 1996; Driver et al. 1998; Marleau \& Simard 1998) and at least five sets of human eyeballs (Abraham et al. 1996; van den Bergh et al. 1996; Driver et al. 1998; Ellis, Abraham \& Dickinson 2001), and the basic result therefore seems quite solid. However only about 5\% of the galaxies in the HDF have known redshifts. Since the deepest existing redshift survey (the Caltech Faint Galaxy Redshift Survey, CFGRS; Cohen et al. 1996) begins to become seriously incomplete around $R=24$ in the HDF ($R=23$ in the HDF flanking fields), existing HST imaging data go significantly deeper than do spectroscopic redshifts. Several photometric redshift-based studies lend support to the passive evolution picture for disks, but there really is nothing like a proper spectroscopic redshift to lend credibility to this kind of work, so it came as something of a relief when Brinchmann et al. (1998) was able to show (on the basis of HST images for 341 galaxies with $I_{AB}<22.5$ mag of known redshift from the CFRS/LDSS redshift survey) that the proportion of  spectroscopically-confirmed disks is indeed in good agreement with passive evolution predictions.

More fine-grained ``precision morphology'' (which can subdivide spiral galaxies into various classes, for example) still requires a trained human eye, and also requires complete redshift information and multi-filter data in order to synchronize data to a uniform rest wavelength.  Using the CFGRS data, van den Bergh et al. 2000 were able to show that the fraction of intermediate/late- type (Sb - Ir) galaxies in the HDF the drops by a factor of $\sim2$ out to $z\sim1$, presumably because many late-type galaxies viewed at a look- back times of $\sim8$ Gyr have been classified as ``peculiar'' or ``merger'', while the fraction of such peculiar or merging galaxies exhibits a monotonic increase from 5\% at $z \sim0$, to 10\% at $z \sim 0.4$, 19\% at $z \sim 0.7$ and 30\% at $z \sim 0.95$. It should be once again emphasized that the numbers of galaxies in this investigation is small, and the conclusions which can drawn from these data are correspondingly uncertain. It is also worth noting that the merger rate estimated from the HDF on the basis of morphological considerations is in reasonable agreement with the merger rate inferred from HST imaging of the CFRS fields, and from pair counts at lower redshifts. 

Beyond $z\sim1$, one must observe galaxies in the infrared in order to study them at familiar optical wavelengths in their rest frames.  The NICMOS camera on HST had this capability, although its small field of view and limited lifetime  meant that relatively few $z>1$ field galaxies were observed in the near infrared before the instrument ceased functioning. However, over the short lifetime of this camera it was possible to demonstrate that most galaxies with peculiar optical morphologies remain peculiar when viewed in the near infrared, at least out to $z=3$ (Dickinson 1999), and as shown in Figure~2 this really ought to be a fair test of the nature of these objects. It follows that the strange appearance of the peculiar objects at these redshifts is not merely the consequence of ``morphological K-corrections" (i.e. of observing only irregularly distributed sites of star formation at rest ultraviolet wavelengths, while missing the bulk of the galaxy). The peculiar appearance of these objects reflects a genuinely irregular structural state in these galaxies. What is unclear, however, is what proportion of these objects are bona fide disks. Furthermore, in the most distant objects (i.e. at $z>3$), it may be the case that we really are only seeing the very tip of the baryonic mass iceberg (i.e. only the youngest stars in the galaxies, grouped together in irregular star formation complexes). 

\begin{figure}
\plotfiddle{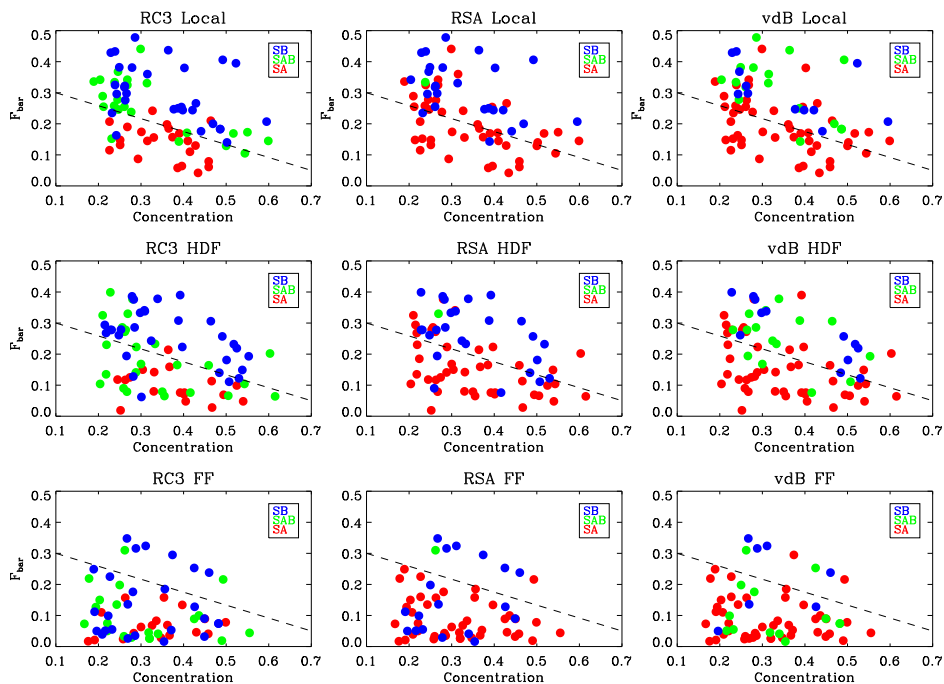} {10cm} {90} {50} {50} {190} {0}
\caption{
An illustration (taken from van den Bergh et al. 2002) of the degradation of bar visibility with redshift and noise, and of observer-to-observer agreement in visual bar classifications. [Top row] The local Hubble space diagram populated with galaxies from the {\em OSU Bright Spiral Galaxy Survey}. Plot symbols keyed to the bar classifications from the RC3 (left panel), the RSA (middle panel), and visual inspection by van den Bergh (right panel), as described in the text. Note the excellent agreement between the bar classifications in the RC3 and position above the dashed line in Hubble space. The middle panel illustrates the well-known fact that the RSA adopts stricter criteria for classification of barred spirals than the RC3 (Binney \& Merrifield 1998). The rightmost panel shows that van den Bergh's visual classifications are closer to the RSA than to the RC3, although van den Bergh classifications allow for more dynamic range within the barred spiral category: many objects classified as S(B) by van den Bergh are classified as SB in the RSA. [Middle Row] As for the top row, except showing the Hubble space diagram for the sample degraded to the conditions in the Northern Hubble Deep Field. Colors are keyed to {\em local} classifications. Strongly barred galaxies are clearly still visible in these data, though weakly barred systems have now migrated to the unbarred portion of the Hubble space diagram. [Bottom Row] As for the previous row, except showing the Hubble space diagram for the sample degraded to the conditions of the HDF Flanking Field observations. Only a few strongly barred spiral galaxies are evident in data of this quality. 
 \label{fig:barvisibility}}
\end{figure}

As described by Merrifield in this volume, a number of studies have also claimed that the percentage of barred spirals seems to decrease towards larger look-back times (van den Bergh et al. 1996; Abraham et al. 1999; van den Bergh et al. 2000). However, an alternative interpretation of the absence of barred and grand design spirals at high redshifts is that the features that define these categories of objects locally are undetectable outside of the nearby Universe. A useful tool in testing this is what Abraham \& Merrifield (2000) have dubbed ``Hubble space'', a quantitative two-parameter description of galactic structure that maps closely on to Hubble's original tuning fork. By  ``artificially redshifting'' samples of local galaxies and mapping them onto Hubble space it is possible to assess the extent to which characteristics such as bar strength should remain visible at high redshifts. Van den Bergh et al. (2002) present such an analysis for a statistically fair set of local spirals taken from  the {\em Ohio State University Bright Spiral Galaxy Survey} (Frogel, Quillen \& Pogge 1996).  Results from this study 
are shown in Figure~\ref{fig:barvisibility}, where it is seen 
%show that 
about $\sim 2/3$ of all luminous local SB galaxies would still classified as SB or S(B) at the resolution and noise level of the Hubble Deep Field, dropping to around 50\% in the flanking fields. Therefore image degradation at high redshift can only account for a fairly small part of the observed decrease in the fraction of SB galaxies with increasing redshift,  although we caution that the total the number of suitably bright (and suitably inclined) galaxies in the HDF data is small and more observations are needed to better establish the putative demise of barred spirals at high redshifts.

Explaining why the Hubble classification scheme appears to ``evaporate'' at redshifts $z > 0.3$ is a challenge that can only be met by a greater understanding of galaxy dynamics at high redshifts. The gas fraction in typical spiral galaxies may have decreased by a large fraction between $z = 1$ and $z = 0$, and encounters with giant molecular clouds, spiral arms and perhaps clumps in the dark matter halo may have increased the velocity dispersion within the disk at high redshifts. Fuchs \& von Linden (1998) suggest that either of these effects would impact the stability of spiral disks in the distant past, but the data simply do not yet exist to allow astronomers to determine whether either of these effects plays a dominant role. At this stage all that is clear is that the morphology of spiral galaxies is evolving rapidly, and systematically, even at quite low redshifts. Familiar types of galaxies, such as barred and grand-design spirals, appear to be relatively recent additions to the extragalactic zoo. The nature of the many morphologically peculiar galaxies at high redshift and their relationship to low-redshift disk systems remains a complete mystery. These objects might be mergers, proto-galaxies, new classes of evolved systems, or a combination of all three. Clearly we are only beginning to glimpse the morphology of galaxies in the distant Universe, and are just at the stage where we are groping to fit them into a physically meaningful organizational structure. 

\section{Conclusions and Speculations}

It seems  that the observable universe falls into two domains. At $z < 1$, 3/4 of luminous galaxies are disks. Only a few of the remaining nearby ellipticals are presently forming stars at a significant rate. So almost all star formation at $z < 1$ is taking place in disks. In these objects star formation is occuring in a manner consistent with passive evolution and with the slow build-up of stellar populations. On the other hand most star formation at higher redshifts  is taking place in mergers, or in rather chaotic objects. Is there a connection between the transition from a merger-driven mode of star formation and the change in the slope of the star-formation history of the Universe plot (the ``Madau plot'') at $z\sim1-1.5$? It is not impossible that the changing mode of star formation is the underlying {\em cause} of the inflection in the Madau plot. If early disks are being destroyed at $z>1$ in this process then one would not expect the Galaxy to have an old metal-poor thin disk. In any case, we are only now entering the phase where optical and near-IR data will be deep enough to allow fair samples of disks at $z>1$ to be studied. The clear expectation is that studies of such objects will show that disk sizes decrease with increasing $z$ because of tidal truncation and destruction, and because gaseous disks are more prone to destruction by the winds induced by violent central bursts of star formation. In any case there is so much evidence that our own Galaxy's disk grew from inside out that it would be surprising if this was not seen in other systems at high redshifts. 

By $z=1$ it appears that most of the big disks have already settled into place, although their morphologies remain outside the formal Hubble sequence. At $z \sim0.7$ about 2/3 of all Sbc-Sc galaxies are peculiar. Clearly late-type disks have mostly not yet had time to revert to their Hubble prototype form. On the other hand, the fraction of peculiar Sa's at $z \sim0.7$ appears to be minute. Some of this is no doubt due to the fact that it might be difficult to see some forms of peculiarity in compact early-type galaxies, but it seems important to determine whether Sa and Sab galaxies at $z \sim0.7$ are basically similar in overall structure to nearby ones, or whether  the nature of spiral structure changes with look-back time in a different way in Sa and Sc galaxies. The best preliminary evidence would seem to suggest that in Sc galaxies the spiral arms become patchier and more chaotic with increasing look-back time. On the other hand spiral arms appear to become less prominent in early-type galaxies at large look-back times. In any case, it is tempting to speculate that some of these effects might be related to the apparent paucity of barred spirals at $z > 0.7$. Perhaps high-redshift disks are too dynamically hot to become unstable in a global bar forming mode.


\begin{references}

\reference Abraham, R.~G. \& Merrifield, M.~R. 2000, AJ, 120, 2835

\reference Abraham, R.~G., Merrifield, M.~R., Ellis, R.~S., Tanvir,
N.~R. \& Brinchmann, J. 1999, MNRAS, 308, 569

\reference Abraham, R.~G.\ 1997, In ``The Ultraviolet Universe at Low and High Redshift: Probing the Progress of Galaxy Evolution'',  AIP Conference Proceedings, v.408., p.195

\reference Abraham, R.~G., Tanvir, N.~R., Santiago, B.~X., Ellis, R.~S., Glazebrook, K., \& van den Bergh, S.\ 1996, \mnras, 279, L47

\reference Bershady, M. 2002, this volume

\reference Bouwens, R., Broadhurst, T., \& Silk, J.\ 1998, \apj, 506, 557

\reference Brinchmann, J.~et al.\ 1998, \apj, 499, 112

\reference Cohen, J.~G., et al. 1996, \apjl, 471, L5

\reference Cowie, L.~L., Hu, E.~M., \& Songaila, A.\ 1995, \aj, 110, 1576

\reference Dickinson, M. 1999. In {\em After the Dark Ages: When Galaxies were Young (the Universe at $2 < z < 5$).}  Edited by S. Holt and E. Smith. American Institute of Physics Press, p. 122.

\reference Dickinson, M.~et al.\ 2000, \apj, 531, 624

\reference Driver, S.~P., Windhorst, R.~A., \& Griffiths, R.~E.\ 1995, \apj, 453, 48

\reference Driver, S.~P., Fernandez-Soto, A., Couch, W.~J., Odewahn, S.~C., Windhorst, R.~A., Phillips, S., Lanzetta, K., \& Yahil, A.\ 1998, \apjl, 496, L93

\reference Driver, S.~P.\ 1999, \apjl, 526, L69

\reference Ellis, R.~S., Abraham, R.~G., \& Dickinson, M.\ 2001, \apj, 551, 111

\reference Faber, S.~M., Phillips, A.~C., Simard, L., Vogt, N.~P., \& Somerville, R.~S.\ 2001, ASP Conf.~Ser.~230: Galaxy Disks and Disk Galaxies, 517

\reference Frogel, J.~A., Quillen, A.~C., \& Pogge, R.~W.\ 1996, ASSL Vol.~209: New Extragalactic Perspectives in the New South Africa, 65

\reference Giallongo, E et al. 2000, ApJ, 530L, 73. 

\reference Griffiths, R.~E.~et al.\ 1994, \apj, 437, 67

\reference Le F{\` e}vre, O.~et al.\ 2000, \mnras, 311, 565

\reference Lilly, S.~J., Le Fevre, O., Crampton, D., Hammer, F., \& Tresse, L.\ 1995, \apj, 455, 50

\reference Lilly, S.~et al.\ 1998, \apj, 500, 75

\reference Marleau, F.~R.~\& Simard, L.\ 1998, \apj, 507, 585

\reference Merrifield M. R. 2002, this volume

%\reference Sandage, A. \& Tammann, G.A. 1981, Revised Shapley-Ames Catalog
%of Bright Galaxies (Washington: Carnegie Institute)

\reference Schade, D., Carlberg, R.~G., Yee, H.~K.~C., Lopez-Cruz, O., \& Ellingson, E.\ 1996, \apjl, 465, L103

\reference Schade, D., Lilly, S.~J., Le Fevre, O., Hammer, F., \& Crampton, D.\ 1996, \apj, 464, 79

\reference Simard, L.~et al.\ 1999, \apj, 519, 563

\reference Steidel, C.~C., Pettini, M., \& Adelberger, K.~L.\ 2001, \apj, 546, 665

\reference van den Bergh, S. et al. 2002. (astro-ph/0202444).

\reference van den Bergh S., Abraham R.G., Ellis R.S., Tanvir N.R. \&
Santiago B.X. 1996, AJ, 112, 359

\reference van den Bergh, S., Cohen, J.~G., \& Crabbe, C.\ 2001, \aj, 122, 611 

\reference van den Bergh, S., Cohen, J.~G., Hogg, D.~W., \& Blandford, R.\ 2000, \aj, 120, 2190

\reference van Dokkum, P.~G.~\& Stanford, S.~A.\ 2001, \apjl, 562, L35

\reference Vogt, N.~P.\ 2001, ASP Conf.~Ser.~240: Gas and Galaxy Evolution, 89

\reference Vogt, N.~P.\ 2000, ASP Conf.~Ser.~197: Dynamics of Galaxies: from the Early Universe to the Present, 435

\reference Vogt, N.~P.~et al.\ 1997, \apjl, 479, L121

\end{references}
\end{document}